\documentclass[12pt]{article}

\usepackage{newtxtext,newtxmath}

\newcommand{\qqq}{\quad\quad\quad}
\newcommand{\abrack}[1]{\left\langle#1\right\rangle}

\usepackage{graphicx}

\usepackage[letterpaper,margin=1in]{geometry}

\renewenvironment{abstract}
	{\quotation}
	{\endquotation}

\date{}

\makeatletter
\renewcommand{\fnum@figure}{\textbf{Figure \thefigure}}
\renewcommand{\fnum@table}{\textbf{Table \thetable}}
\makeatother

\usepackage{scicite}

\usepackage{url}

\def\scititle{
	Cross-feeding Creates Tipping Points in Microbiome Diversity
}
\title{\bfseries \boldmath \scititle}

\author{
	Tom~Clegg$^{1,2\ast}$,
	Thilo~Gross$^{1,2,3}$. \and 
    \small$^{1}$Helmholtz Institute for Functional Marine Biodiversity at the University of Oldenburg, Germany.\and
    \small$^{2}$Alfred Wegner institute, Bremerhaven, Germany.\and
     \small$^{3}$ICBM, University of Oldenburg, Oldenburg, Germany.\and
	\small$^\ast$Corresponding author. Email: thomas.clegg@hifmb.de\and
}

\begin{document} 

\maketitle

\begin{abstract} \bfseries \boldmath
A key unresolved question in microbial ecology is how the extraordinary diversity of microbiomes emerges from the behaviour of individual populations. This process is driven by the cross-feeding networks that structure these communities, but are hard to untangle due to their inherent complexity. We address this problem using the tools of network science to develop a model of microbial community structure. We discover tipping points at which diversity abruptly declines due to the catastrophic collapse of cross-feeding networks. Our results are a rare example of an ecological tipping point in diversity and provide insight into the fundamental processes shaping microbiota and their robustness. We illustrate this by showing how the unculturability of microbial diversity emerges as an inherent property of their microbial cross-feeding networks. 

\end{abstract}

\clearpage

\noindent
Microbiomes are amongst the most diverse ecological systems on Earth, consisting of hundreds of populations interacting in complex networks of resource consumption and exchange \cite{Horner-Devine2004,Thompson2017,Faust2012}. They are ubiquitous and perform many vital functions from the cycling of nutrients in ecosystems \cite{Schimel2012} to the mediation of gut health \cite{Hou2022}. As advances in sequencing technologies allow us to examine these systems in increasing detail, there is a growing interest in answering fundamental questions about the ecology of microbiota and how their extraordinary diversity is maintained \cite{Antwis2017,eren2024modern}. Of particular importance are their cross-feeding dynamics in which microbial consumers secrete the by-products of their metabolism into the environment allowing them to be used by other members of the community \cite{DSouza2018}. Cross-feeding has been shown to be widespread, helping to maintain both the high diversity and varied functional capacity of microbiota \cite{Goldford2018, Kost2023}.

Although microbial communities are important, untangling their inherent complexity is difficult. In response, a recent surge of work has successfully drawn on methods from theoretical ecology to address this problem \cite{Coyte2015,Butler2018,mentges2019long,Marsland2019,Grilli2020,clegg2024}. Much of this is built on the insight that community-level properties are emergent \cite{Andersen1972}, arising from the pattern of interactions between populations rather than any individual component. This approach allows us to link the structure of interaction networks to the behaviour of complex ecological systems \cite{Montoya2006, Williams2000,Allesina2008,yeakel2020}, and provides a powerful tool to address the complexity of microbiota and their cross-feeding networks.

Viewing the microbiota through the lens of ecological theory also opens up new questions. The stability of complex communities has been a topic of persistent importance for over a century \cite{May1972,Hastings1982,Allesina2015,Gross2009}. One question that has been raised is whether there are systemic tipping points in these systems at which small changes in structure results in abrupt, discontinuous changes in community-level properties such as diversity \cite{Scheffer2001}. Examples of transitions in simple systems such as the well-known clear-turbid transition in lakes are abundant \cite{Scheffer1993}, but whether they can exist in complex, diverse communities like microbiota is debated, and important for our understanding of biodiversity loss \cite{Kefi2022,Hillebrand2023}. 

Although identifying tipping points in complex systems such as microbiota can be challenging, one way we can approach problem is with the tools of network science \cite{newman2003structure}. From a networks perspective, microbes and metabolites can be described as a bipartite graph or hypergraph, in which consumer populations are linked by directed hyperedges representing the metabolites they consume and secrete. Doing so allows for the analyses with techniques from  percolation theory for higher-order networks which are uniquely suited for the identification of structural transitions \cite{Callaway2000,Buldyrev2010,Gao2012,Niu2016}. Their application led to the important finding that these types of tipping points tend to be rare and there are very few examples of such behaviour \cite{Dsouza2019}. They are so rare that the discovery of explosive percolation was a major breakthrough \cite{Achlioptas2009}, although it was later shown that this transition is technically not discontinuous \cite{Riordan2011}. 

In this paper, we apply methods from network science to understand how cross-feeding networks mediate the maintenance of in diversity in complex microbiomes. We formulate a simple, generic model of a microbial community, inspired by recent work \cite{mentges2019long,Marsland2019}, that captures the cross-feeding dynamics between consumer populations. The interactions between populations via the exchange of metabolites leads naturally to a network representation of the community structure, which we analyse utilising tools from percolation theory on higher-order networks. We show how the persistence of the consumers within the community can be defined in terms of simple statistical features of their cross-feeding network, letting us link the structure of interactions to the emergent diversity. Furthermore, we show that the model has abrupt, discontinuous transitions corresponding to a rare example of a ecological tipping point between states of high and low community diversity. Finally we show how this tipping point can provide insight into the behaviour of real microbiota by considering how the culturable diversity in communities emerges from the structure of their interaction networks and their robustness to perturbation.  

\subsection*{The Microbial Community Hypergraph model}

\begin{figure}
    \centering
    \includegraphics{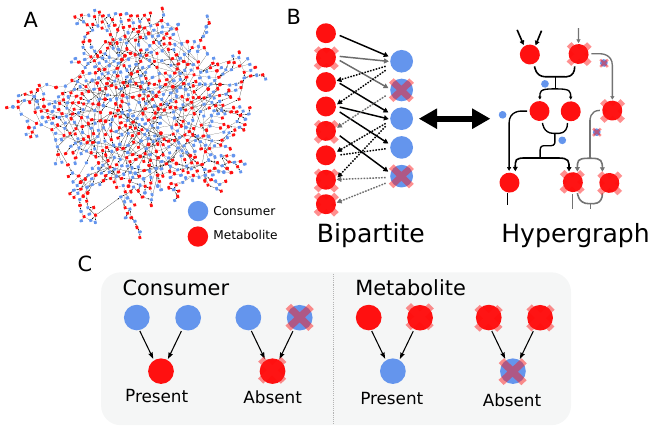}
    \caption{\textbf{Cross-feeding networks and the microbial community model}. Diagram showing the overall structure of the microbial community model. (A) A microbial community with a complex random cross-feeding network. Nodes represent consumers (blue) and metabolites (red) with links indicating the resource requirements $M \rightarrow C$ and metabolite production $C \rightarrow M$. (B) The network has alternative representations as a bipartite graph or a hypergraph where consumer nodes are linked by directed hyperedges representing metabolite uptake and release. Consumers and metabolites are marked present or absent (nodes with a cross) based on the rules below. (C) Simple rules dictate the presence of populations and metabolites. Consumers are present when all their required metabolites are present (left). Metabolites are present when any one of the producers is present (right).}
    \label{fig:ModelDiagram}
\end{figure} 

We consider a complex community of $N$ microbial populations interacting via the consumption and exchange of a set of $M$ metabolites (Fig.~\ref{fig:ModelDiagram}). Following \cite{mentges2019long}, we assume that each population requires a fixed set of metabolites to persist in the system. As a result of their metabolic activity, each population also produces a set of metabolites, which is released by continuous secretion or upon the death of the cell. The model can thus be imagined as a bipartite network of populations and metabolites, or alternatively as a hypergraph where the populations are the nodes and the metabolites that connect them are directed hyperedges (Fig.~\ref{fig:ModelDiagram}B). 

For simplicity we do not model the dynamics of the system and focus entirely on structural feasibility, focusing on communities structured by cooperative cross-feeding interactions as opposed to competition. Each population and metabolite is described by a binary presence/absence variable (Fig.~\ref{fig:ModelDiagram}C). A metabolite is present if and only if, at least one population that produces the metabolite is present. Conversely a population is present if, and only if, all metabolites required by the population are present.

We consider a statistical ensemble of such models in which the number of metabolites required by populations and the number of producers of each metabolite are drawn from the probability distributions $c_k$ and $m_k$ such that, for example, $m_3$ is the probability that a metabolite is produced by 3 populations. By selecting $c_k$ and $m_k$ appropriately we can emulate the desired features of a microbial community such as the relative numbers of currency metabolites (i.e. those required by many populations) or common waste products. We study the full ensemble of networks consistent with these distributions.

Below we show that the model undergoes a structural transition, reminiscent of the appearance of the giant component in simple graphs \cite{Molloy1995}. We analyse this transition by building on an elegant formalism \cite{Newman2001}, which uses mathematical objects called generating functions to store and manipulate sequences of numbers (Supplementary Text). These generating functions capture the structure of the network by encoding the distributions of the number of consumer requirements and metabolite producers. For the present hypergraph model the generating functions are defined as 
\begin{equation}
    \label{eq:PGF}
    C(x)=\sum c_k x^k \qqq M(x)=\sum m_k x^k,   
\end{equation}
where $C(x)$ encodes the numbers of resource requirements across consumers, $M(x)$ the number of producers of each metabolite and, $x$ is an abstract variable introduced for technical reasons (Supplementary Text). 

Using these generating function we can determine the proportions of populations $c^*$ and metabolites $m^*$ that persist in the community. In ecological terms these represent the relative diversity of the system. We start by considering the conditions for persistence, that consumers must have all their requirements met but metabolites need only one producer. We can express these probabilistically using the generating functions in Eq.~\ref{eq:PGF} (Supplementary Text), leading to
\begin{align}
    \label{eq:prop_present}
    \begin{split}
            c^* &= C(m^*),  \\
            m^* &= 1 - M(1 - c^*). 
    \end{split}
\end{align}
From Eq.~\ref{eq:prop_present} we can solve for $c^*$ and $m^*$, giving the stable configurations of communities in which the needs of consumers are met through the cross-feeding network. Though it appears simple, this expression captures the complexity of the interdependencies between populations and how consumers rely on the populations they interact with who in turn depend on their own interacting partners and so on. Crucially, it does so in terms of the statistical properties of the network, allowing us to link the structure of interactions between populations to the emergent diversity the community cross-feeding network can support. 

\subsection*{Structural Tipping Points and Diversity}

\begin{figure}
    \centering
    \includegraphics{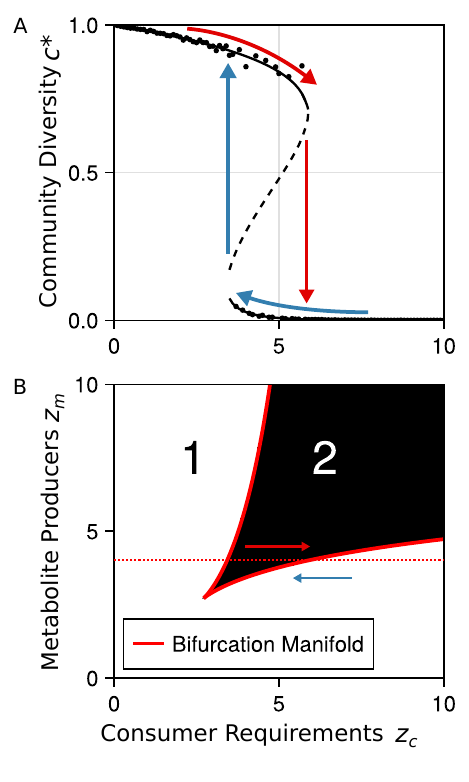}
    \caption{\textbf{Cross feeding networks have tipping points in consumer diversity}. (A) Bifurcation plot showing the change in diversity as consumer requirements $z_c$ varies and $z_m = 4$. Lines show the predicted values from Eq.~\ref{eq:prop_ER}, with solid lines indicating stable states and dashed lines unstable. Each point is from a randomly generated network with given structural properties with $N=M=10000$. Arrows indicate the path-dependence in the system. Increasing $z_c$ (red arrows) leads us over the tipping point from a high to low diversity state whist decreasing (blue arrows) does the opposite. (B) Phase plot showing the arrangement of tipping points and diversity in the 2D-parameter space. Changes in diversity are continuous in the white area (1). The points at which transitions happen form the bifurcation manifolds (red lines). The shaded area (2) indicates the region of bistability and path dependence where high and low diversity states are possible depending on the history of the system. The dotted red line and arrows indicate the slice over which figure (A) is displayed.}
    \label{fig:bifurcation}
\end{figure} 

To explore the patterns of diversity and community structure we consider random cross-feeding networks in which metabolite requirements and production are distributed across the community. These random hypergraphs are reminiscent of Erdős–Rényi networks \cite{Erdos:1959}, but produce some surprising results due to the structure imposed by our microbial community model. Although they do not include all the biological realism, these types of networks are useful as they help us build intuition about the effects of structure and tend to be reasonable approximations in large complex systems \cite{Newman2002}.

In random networks the numbers of metabolite requirements and producers follow a Poisson distribution with generating functions $C(x) = \exp(z_c (x-1)), M(x) = \exp(z_m (x-1))$ where $z_c$ and $z_m$ are the average number of resource requirements and metabolite producers respectively. Applying these to Eq.~\ref{eq:prop_present} lets us write the self-consistency equation 
\begin{equation}
    \label{eq:prop_ER}
    c^* = \exp(-z_c \exp( -z_m c^*)).
\end{equation}
Solving Eq.~\ref{eq:prop_ER} for $c^*$ reveals how the stable level of consumer diversity varies depending on the distribution of consumer requirements and metabolite producers characterised by $z_c$ and $z_m$ (Fig.~\ref{fig:bifurcation}, Fig~\ref{fig:si:graphical}). Overall, community diversity increases as a function of the average number of metabolite producers $z_m$ but falls with increasing consumer requirements $z_c$. This relationship makes sense, more consumer populations are able to persist when their requirements are more likely to be produced or when they have fewer requirements in the first place. 

The change in diversity with network structure is initially continuous. However, if one varies consumer requirements $z_c$ with a sufficiently high number of metabolite producers $z_m$ or vice versa, discontinuous transitions become possible. For example, if we fix $z_m$ to 4 and slowly increase consumer requirements $z_c$ from 0, community diversity gradually falls until a critical point is encountered (known as fold bifurcations in dynamic systems theory) and the system enters a low diversity state (Fig.~\ref{fig:bifurcation}A). This transition is driven by the collapse of the cross-feeding network, increased requirements result in the loss of consumers which causes the loss of the populations reliant on their metabolites and so on. Conversely, if we take a community and reduce $z_c$ from above it will remain stuck in the low diversity regime, even as it passes the original tipping point, until a second critical point is reached and the community jumps back to the high diversity state. In this case reductions in requirements allow the persistence of populations who produce metabolites others need, allowing the cross-feeding network to reassemble. Between these two transition points is a region of path-dependence or hysteresis, where the system can be either in a state of high or low diversity, depending on its history. 

Considering variation in both consumer requirements $z_c$ and metabolite producers $z_m$ simultaneously results in a two-dimensional parameter space in which transitions can occur (Fig.~\ref{fig:bifurcation}B). In this space the transition points form two curves marking the points at which the high and low-diversity states appear, also known as bifurcation manifolds. These curves eventually meet at a single point $z_m = z_c = e$ and annihilate in a cusp bifurcation which acts as an organising centre for the tipping point behaviour.

Both the bifurcation manifolds and the cusp point can be derived analytically and agree with numerical solutions to Eq.~\ref{eq:prop_ER}. These results are robust to additional structural features such as correlations in network structure (Supplementary Text; Fig.~\ref{fig:cor_phase} and \ref{fig:cor_div}). We also validated these predictions with simulated microbial cross-feeding networks which showed near perfect agreement with the predictions including the expected patterns in diversity and tipping point behaviour (see Material and Methods).

\subsection*{Community Robustness and the Culture of Microbial Diversity}

\begin{figure}
    \centering
    \makebox[\textwidth]{\includegraphics[width=12.1cm]{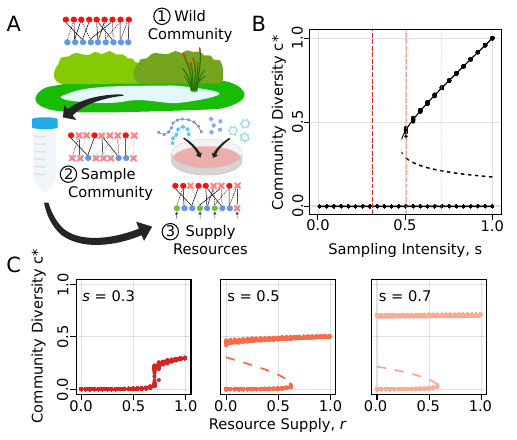}}
    \caption{\textbf{Cultured community diversity collapses due to tipping points in cross-feeding networks}. (A) Diagram showing how sampling a wild community and culturing on a given media affects cross-feeding network structure in the model. From a wild community (1), sampling removes some proportion of consumer populations and any metabolite they produce (2). Supplying resources in culture allows the partial recovery of the cross feeding network (3) giving the final community. (B) Plot showing how community diversity $c^*$ increases as a function of the intensity of sampling $s$. Lines show analytical solutions from Eq.~\ref{eq:prop_ER}, solid lines are stable and dashed unstable. Points show solutions obtained from randomly generated networks with $N=M=1000$ and 100 replicates at each value of $s$ (or $r$ below). Dashed lines indicate the values of $s$ in the plots below. (C) Plots showing the final community diversity $c^*$ against resource supply $r$ across three levels of $s$. Again lines and points show solutions from analytical and numerical simulations respectively.}
    \label{fig:Sampling}
\end{figure}

To illustrate the tipping point and further explore the robustness of community diversity to changes in network structure we next apply our model to ask why are we unable to culture so much of the microbial diversity we see in nature?

This question has long been discussed in the literature, with one common explanation being the obligate dependence of many microbes on cross-feeding interactions \cite{Pande2017}. This idea is supported by evidence for widespread auxotrophies in microbes (the inability to produce essential metabolites required for growth) as well as work showing the ability to grow otherwise unculturable strains using co-culture techniques \cite{lewis2021innovations, Kost2023}. 

We approach this problem by considering the how the culturing of microbial communities acts as a perturbation to their cross-feeding networks and the diversity they are able to support (Fig.~\ref{fig:Sampling}A). We consider this process using our model in two steps, in which we first sample a community from the environment, collecting only a proportion of the populations within, and then culture it in media, providing it some of the resources it needs to grow. This procedure not only acts as a direct perturbation to communities, removing consumers and supplying resources, but also has secondary effects as the loss and gain of consumers and metabolites propagates the cross feeding network. 

While enumerating the effects of these perturbations on individual populations is hard, the use of generating functions makes this problem simple. In our model we can encode the sampling and the supply resources in their own functions $S(x) = 1 + s(x - 1)$ and $R(x) = 1 + r(1 - x)$ where $s$ and $r$ are the proportion of consumers captured in the sample and the proportion of resources supplied respectively. Combined with Eq.~\ref{eq:PGF} these let us express the structure of the cross-feeding network (subscript ${\rm A}$) after sampling and culture succinctly in a new set of generating functions (see Supplementary Text)
\begin{align}
\label{eq:PGF_a}
    C_{\rm A}(x) = C(R(x)) \qqq M_{\rm A}(x) = M(S(x)).
\end{align}
These generating functions capture the probabilities that a consumer or resource has a given number of requirements or producers in the final community after culture, accounting for the change in the average number of consumer requirements $y_c = (1-r) z_c$ and metabolite producers $y_m = (1-s) z_m$ due to the loss and gain of consumers and resources. These can then be used to solve for the proportion populations and metabolites present similar to Eq.~\ref{eq:prop_present}. 

Fig.~\ref{fig:Sampling}B and C show how the process of sampling and culture leads to the diversity of the final cultured community. In the first stage sampling a smaller proportion $s$ of the wild community results in both reduced diversity and fewer metabolite producers $y_m$. If our sampling is incomplete it is possible that too many producers are lost and the system is pushed past the tipping point as the cross-feeding network breaks down into the low diversity state. The subsequent supply of resources in culture effectively reduces the number of metabolite requirements of consumers $y_c$, increasing the persistence of consumers and the relative diversity in the community. Depending on structure of the system, the supply of resources may allow communities to persist in the high diversity state. However, due to the path-dependency discussed above, reassembling the community may be difficult and a greater proportion of resources will need to be supplied to counter the loss of metabolite producers via sampling. 

Together these results provide an explanation for why it is so hard to culture much of the microbial diversity we see in nature, that the inherent structure of their cross-feeding networks creates tipping points that make them fragile to perturbation. If we sample insufficiently or don't supply enough of the resource requirements it is unlikely that the community cross-feeding network will maintain its function, ultimately leading to the loss of diversity.

\subsection*{Conclusion}
In this paper we developed a model of microbial community structure, addressing the question of how the diversity of complex microbiomes emerge from the interactions of individual populations. By utilising tools from network science, our results link the exchange and consumption of metabolites in cross-feeding networks to the emergent diversity and robustness of microbial communities. Importantly, our work reveals the existence of tipping points at which communities can jump between states of high and low diversity as cross-feeding networks collapse and reassemble.

The observation of these behaviours in such a simple model suggests they may be generally applicable in systems structured in this way. By using techniques from network science we are able to link cross-feeding interactions to emergent community diversity. We expect this mechanism to be widely applicable due to the ubiquitous nature of auxotrophies and obligate cross-feeding \cite{DSouza2018,Kost2023}. 

Though we do not consider other types of interactions such as competition between consumers these could in theory be incorporated into our model. In general we expect the transitions we observe to persist but it would be interesting to see if if their behaviour is altered or new tipping points in diversity appear. We propose that future work should continue to use this structural approach to develop models of microbial communicates, especially to include a diversity of interaction types. 

Our results also stand as a rare example of a tipping point in a diverse, complex ecological system. There is an ongoing debate on the question of whether these types of systems undergo tipping points and to what degree they are a useful concept in understanding biodiversity decline \cite{Kefi2022,Hillebrand2023}. Our results contribute to this discussion by demonstrating a well-defined tipping point in diversity and how it can be reached through disturbance of the community. Identifying how the structural mechanisms we identify here relate to the robustness of ecological systems more generally is an exciting line of future research and will hopefully contribute our understanding of diversity loss more generally.

The tipping points and the path-dependent behaviour we observe suggest a inherent fragility in the structure of microbiota. We find that perturbations can cause the collapse of cross-feeding networks and their associated diversity and that recovery may be difficult due to the interdependencies they create between populations. These results provide a testable prediction, that perturbations such as the removal of populations should result in abrupt changes to diversity, as illustrated in our example of the sampling and culture of communities. Experiments to test this should, in principle, be relatively easy to carry out and combined with our results have the potential to improve our understanding of microbiota and their robustness.

\clearpage


\clearpage

\clearpage 

%
\bibliography{export} 
\bibliographystyle{sciencemag}








\newpage


\renewcommand{\thefigure}{S\arabic{figure}}
\renewcommand{\thetable}{S\arabic{table}}
\renewcommand{\theequation}{S\arabic{equation}}
\renewcommand{\thepage}{S\arabic{page}}
\setcounter{figure}{0}
\setcounter{table}{0}
\setcounter{equation}{0}
\setcounter{page}{1} 







\subsection*{Materials and Methods}
\paragraph{Simulating Microbial Communities}
In order to verify the analytical predictions of Eq.~\ref{eq:prop_ER} we generated random microbial cross-feeding networks and demined the proportion of persisting consumer diversity based on the rules discussed in the main text. 

To generate random network structures with defined distributions of consumer requirements and metabolite producers we use a configuration model type approach \cite{newman2003structure}. In these networks we consider directed links to match the flow of resources, i.e. consumption is represented by links from metabolites to consumers and metabolite secretion from consumers to metabolites. 

We start by initialising the system with $N$ consumer populations and $M$ metabolites. We then draw their respective number of requirements and producers from Poisson distributions with the average values $z_c$ and $z_m$, thus defining the endpoints of each link in the network. In the supplementary text we show how constraint that the number of start- and endpoints of links must be equal across the system means that the average value of the in- and outdegree are proportional such that $z_m^{\rm in} = a z_c^{\rm out}$ where $a = M/N$ (i.e. the average number of producers of each metabolite is proportional to the average number of metabolites consumers produce). The converse relation must also hold  $z_c^{\rm in} = a z_m^{\rm out}$ for requirement links. Thus, we sample the start points of each link from a Poisson distribution conditioned on their sum, which we obtain by summing the endpoints in the first step. The distribution of start points thus has a multinomial distribution allowing simple and efficient generation of feasible node-degree distributions.  

Once we have a set of consumer requirements and metabolite links we randomly wire the network. We first we convert the in and out requirement and production sequences into their run-length encoding. This converts the sequence of the number of links for each component into the index to which each link belongs. For example, three consumers with 1, 3 and 2 needs respectively have the degree sequence $[1,3,2]$ which has the run-length encoding $[1,2,2,2,3,3]$. To assign links we then simply shuffle the respective start- and endpoints for the requirement and production links and create links between pairs in the same positions. 

Once we have a network structure with the desired degree distribution we determine the proportion of consumers persisting with a simple iterative approach. We initialise the system by marking a random proportion of consumers and metabolites present. We then move through all nodes in the network in random order, updating their presence based on the rules discussed in the main text. After every loop we check to see if the state of the system has changed (i.e. has any node changed state). Over several iterations the system converges to a steady state in which the only consumers present are those whose needs are met by the metabolites available and vice versa. We stop the process once the system state stops changing. The proportion of persisting consumers and metabolites can be calculated as the ratio of the number present to the total number in the network.

\clearpage
\subsection*{Supplementary Text}

\subsubsection*{Generating Functions}
Generating functions (GF) are mathematical objects that allow us to represent sequences of numbers as coefficients in a power series. GFs are invaluable when working with complex networks as they allow us to deal with the combinatorial explosions that often follow when we try to enumerate though the many possible connections in a network. In this section we will provide a basic overview of their properties and application (for a more complete review for their application to networks see \cite{Gross2022}).

As seen in the main text a generating function for the distribution of a discrete random variable $a$ is defined as
\begin{equation}
    G(x) = \sum p_k x^k
\end{equation}
where $x$ is an arbitrary complex variable and $p_k$ is the probability mass at the value $k$ which must sum to one over the support of $a \in \mathcal{A}$ such that $\sum_{k \in \mathcal{A}} p_a(k) = 1$. In terms of the generating function this is expressed as
\begin{align}
    G(1) = \sum p_k = 1
\end{align}
which is called its norm. We can also recover the point probabilities from a generating function by taking the correct terms of the series expansion
\begin{align}
    p_k = \frac{G^{(k)}(0)}{k!}.
\end{align}
Calculating the moments of the distribution is simple too, for example to calculate the mean we simply take the first derivative and evaluate at $1$
\begin{align}
    G'(1) = \sum_k p_k k = \abrack{k}.
\end{align}
This can also be generalised to higher moments
\begin{align}
    \abrack{k^n} = \sum_k p_a(k) k^n = (k G'(1))^{(n)}
\end{align}
 where we iteratively take the derivative and multiply by $k$ $n$ times.

Some of the most powerful features of generating functions are the operations we can preform on them. For example they can be used to easily calculate the sum of two discrete random variables. To illustrate this it is useful to consider an example where we we have two identical four sided dice and we want to compute the distribution of their sums. The generating function for each die is identical and can be written as
\begin{align}
    G_{\rm die}(x) = p_1 x + p_2 x^2 + p_3 x^3 + p_4 x^4
\end{align}
We can first attempt the calculation manually, enumerating the ways we can arrive at each total sum. For example to arrive at the value of $4$ we can either roll two $2$s or a $1$ then a $3$ or a $3$ then a $1$ giving a total probability of $p_2 p_2 + 2p_1p_3$. Writing out the full equation quickly realise that the coefficients follow the combinatorics of polynomials and that we can write the generating function for the sum of the two die as
\begin{align}
    G_{\rm 2 die} = \left(p_1 x + p_2 x^2 + p_3 x^3 + p_4 x^4 \right)^2 = G_{\rm die}(x)^2.
\end{align}
Thus, the generating function of the sum of any two random variables can be calculated simply multiplying their respective generating functions, avoiding the need for cumbersome enumeration through all possible values it can take.

Generating functions also allow the enumeration of sequential events in the "dice of dice" rule which is useful when considering network structure. For example imagine a scenario in which we want to roll our 4-sided dice to determine how many times we flip a coin. What is the distribution for the total number of heads? To address this we can first write a generating function for our coin denoting tails as $0$s and heads as $1$s
\begin{align}
    G_{\rm coin}(x)= c_0 + c_1 x.
\end{align}
Again, we can try to write the generating function for the total manually, enumerating through all possible combinations. First we consider the case where you get a one on the die roll with probability $p_1$, this means we get one coin toss, which is generated by $G_{\rm coin}$. With a probability $p_2$ we roll a two and get two coin tosses, but applying the rule above we know that the sum of two coin tosses is generated by $(G_{\rm coin})^2$. The same rule can be used for rolls of 3 and 4, and putting this all together gives us the generating function 
\begin{align}
    \begin{split}
        G_{\rm game}(x) &= p_1 G_{\rm coin}(x) + p_2 {G_{\rm coin}}^2(x) + p_3 {G_{\rm coin}}^3(x)  + p_4 {G_{\rm coin}}^4(x) \\  
        &= G_{\rm die}(G_{\rm coin}(x))
    \end{split}
\end{align}
The second line shows that the result of this game can be generated by applying the generating function of the die to the generating function of the coin toss. This is the dice-of-dice rule: If we are summing over a random number random events then the generating function for the outcomes is the generating function of the events plugged into the generating function for the number of these events. 

\subsubsection*{The Microbial Hypergraph Model}

In this section we show the derivation of the microbial hypergraph model in more detail, discussing the discontinuous percolation transition and extensions of the model to correlated consumer and metabolite degree. 

As discussed in the main text we consider a community of $N$ microbial populations and $M$ metabolites. Each consumer has a set of resources that they require to persist, the number and identity of which varies amongst species. A consumer population is able to persist if all the resources it requires are present. It will use these resources to fuel growth and reproduction creating metabolic by-products in the process. Metabolites are present in the system if any of the consumer populations that produce them are present. 

The model above has multiple interpretations as a network. We can consider the system a bipartite graph where the two components, populations and metabolites, are connected by two sets of directed links representing the requirements and production of resources respectively. Alternatively we can consider a directed hypergraph representation where the consumer nodes are linked via their connections to specific metabolites. In this framework each metabolite represents a directed hyperedge linking the producers of the metabolite to its consumers. Following the rules describe above, nodes in the hypergraph will be active only if all incoming hyperedges are present but edges can be activated by any source node. 

The microbial communities model can be investigated using the mathematics of generating functions discussed above. We first define two probability generating functions to describe the distribution of populations requirements and metabolite producers
\begin{equation}
    \label{eq:si_generatingfunctions}
     C(x)=\sum c_k x^k \qqq M(x)=\sum m_k x^k   
\end{equation}
where $C(x)$ and $M(x)$ are the generating functions for the numbers of consumer requirements and metabolite producers and $c_k$ and $m_k$ are the probabilities that a consumer or metabolite have indegree $k$ respectively. Following the standard definitions of probability generating functions we can calculate important quantities such as the average number of requirements and metabolite producers by taking the derivative and evaluating at $1$
\begin{align*}
    z_c = C'(1) \qqq z_m = M'(1).
\end{align*}

\paragraph*{Community diversity}
We now ask what is the number of consumer populations and metabolites that can be supported in a community with a given distribution of metabolite requirements and producers (i.e. $C(x)$ and $M(x)$). Denoting the proportion of populations and metabolites present as $c^*$ and $m^*$ we first determine the probability that a randomly selected consumer population is present. This is the same as asking what is the chance we select a consumer with $k$ requirements and that those $k$ metabolites are also present leading to 
\begin{align}
    \label{eq:si_c_prop}
    c^* &= \sum c_k (m^*)^k.
\end{align}
We can also apply the same approach to the metabolites, asking what is the chance we select a metabolite with $k$ producers and that at least one of these is present
\begin{align}
    m^* &= \sum m_k (1 - (1 - c^*)^k) \nonumber \\
    &= 1 - \sum m_k (1-c^*)^k, \label{eq:si_m_prop}
\end{align}
where we have used the fact that the probabilities sum to 1, $\sum m_k = 1$ in the last step.

We can see that equations~\ref{eq:si_c_prop}~+~\ref{eq:si_m_prop} both have forms identical to the generating functions defined in Eq.~\ref{eq:si_generatingfunctions}. This lets us write the system in terms of the generating functions leading to 
\begin{align}
    \label{eq:si_sys_eq}
    \begin{split}
        c^* &= C(m^*),\\
        m^* &= 1 - M(1 - c^*).
    \end{split}
\end{align}
From this set of equations it is easy to exclude $m^*$ and obtain a self-consistency equation
\begin{equation}
    \label{eq:si_sc}
    c^* = C(1 - M(1-c^*)),
\end{equation}
which can be solved for $c^*$. The solutions to equation \ref{eq:si_sc} give the proportion of total consumer populations that can persist and thus the relative diversity in the community. With the value for $c^*$ we can also substitute back into Eq.~\ref{eq:si_sys_eq} and obtain the corresponding values of $m^*$.

\subsubsection*{Random Networks}
In this section we consider what proportion of the community is able to persist in a random network. In a random network populations require each metabolite with a fixed probability $p_c$ and metabolites are produced by each consumer with a probability $p_m$. In the limit of a large network where $N,M \to \infty$ the numbers of consumer requirements and metabolite producers approach a Poisson distribution with parameters $z_c = M p_c$ and $z_m = N p_m$, the average requirement and production degree. As they are Poisson distribution the generating functions $C(x)$ and $M(x)$ have the simple form
\begin{equation}
    \label{eq:si_pgf}
    G(x) = \exp[z (x - 1)]
\end{equation}
where $z$ is the average degree. Using Eq.~\ref{eq:si_pgf} as the generating functions for population requirements $C(x)$ and metabolite production $M(x)$ yields 
\begin{equation}
    \label{eq:si_rand_self}
    c^* = \exp\left[-z_c \exp\left(-z_m c^* \right) \right]
\end{equation}
which can be solved to give the solutions for the proportion of consumer population present in the system $c^*$. These can in turn be used to solve for the metabolites $m^*$. 

\subsubsection*{Tipping Points and Bifurcation analysis}
We now consider how the solutions to Eq.~\ref{eq:si_rand_self} give rise to tipping points in community diversity as the structure of cross-feeding network changes. Numerical solutions show that diversity within the community $c^*$ undergoes a bifurcation once the average number of consumer requirements $z_c$ and metabolite producers $z_m$ pass as certain point (Fig.~\ref{fig:bifurcation}). We can derive the points at which these transitions occur by determining the region of parameter space at which the solutions to Eq.~\ref{eq:si_rand_self} also have a zero-valued derivative. 

We start by taking the derivative of Eq.~\ref{eq:si_rand_self} and setting it to zero
\begin{equation}
    \label{eq:si_deriv_c}
    0 = z_c z_m \exp(-z_c \exp(-z_m c^*) - z_m c^*) - 1.
\end{equation}
Values of $z_c$, $z_m$ and $c^*$ that satisfy this equation and Eq.~\ref{eq:si_rand_self} thus correspond to critical points where the bifurcation occurs. We next solve Eq.~\ref{eq:si_deriv_c} to get an expression for the $c^*$ in terms of the two structural parameters
\begin{equation}
    c^* = -\frac{W(-\frac{1}{z_c})}{z_m}
\end{equation}
where $W(x)$ is the Lambert W function which gives the solution to
equations of the form $y = xe^x$ as $x = W(y)$. For real values (such as the probabilities we consider here) the Lambert W function has two branches over the range $-1/e < x < 0$. Substituting this solution for $c^*$ back into Eq.~\ref{eq:si_rand_self} gives an equation for the bifurcation manifold in terms of the structural parameters
\begin{equation}
    \label{eq:si_bifurcation}
    z_m = -\frac{W(-1/z_c)}{\exp(W(-1/z_c) ^ {-1})}.
\end{equation}
This expression gives the curve that marks the point of the discontinuous transition in the model. It is made of up of two branches which mark the transitions between high and low diversity states in either direction. Fig.~\ref{fig:bifurcation}B shows the arrangement of the manifolds, including the region in-between where multiple community states are possible depending on the history (i.e path dependency) in the system.

We can also use Eq.~\ref{eq:si_bifurcation} to consider the point at which the transition originates. As the Lambert W function is defined over the range when its argument $x > -1/e$. This means that the smallest value of $z_c$ for which the bifurcation can occur is at $z_c = e$ which in turn gives the minimum value of $z_m = e$. Thus the cusp bifurcation, where the two manifolds marking the tipping points collide is at $z_c = z_m = e$ 

\subsubsection*{Correlated Degrees}
In this section we consider the extension of the model with correlations between the in- and outdegree of consumer populations and metabolites. These correlations may arise if, for example, consumers that have many resource requirements also tend to produce a variety of metabolites (i.e. they have high metabolic diversity). Likewise metabolites that are produced by many populations may also tend to be used by many consumers representing the tendency of consumers to use resources available to them in the environment. 

In order to include correlations we need to consider the joint in/outdegree distributions of nodes by using bivarate generating functions
\begin{align}
    C(x,y) &= \sum_{jk} c_{jk} x^j y^k, \\
    M(x,y) &= \sum_{jk} m_{jk} x^j y^k.
\end{align}
These bivarate generating function follow the same basic principle as their univarate counterparts, representing the joint distribution of in- $j$ and outdegrees $k$. The probabilities $c_{jk}$ and $m_{jk}$ encode the correlation between in- and outdegree for consumers and metabolites. The direction of edges follow the flow of material, metabolite consumption is represented by edges flowing out of metabolites and into consumers and vice versa.

We can derive a number of quantities from the generating functions including the marginal marginal distribution of in or out links
\begin{align}
    \label{eq:si:marginal}
    \begin{split}
          C^{\rm in}(x) &= \sum_{jk} c_{jk} x^j 1^k = C(x,1),\\
          C^{\rm out}(x) &= \sum_{jk} c_{jk} 1^j y^k = C(1,y),
    \end{split}
\end{align}
as well as the marginal averages
\begin{align}
        z^{\rm in}_c &= \sum_{jk} c_{jk} j = 
        \left. \frac{\delta C(x,y)}{\delta x} \right|_{x,y=1} \\
\end{align}

It is also important to note that as we now explicitly track both in- and outdegree of each node we need to make sure the total degree across the network adds up. This places a constraint on the degree distributions and can be expressed in the equalities
\begin{align}
\begin{split}
    \sum^N_{i} j_i = \sum^M_{a} k_a  \\
    \sum^N_{i} k_i = \sum^M_{a} j_a
\end{split}
\end{align}
where $j_i$ is the indegree of the $i^{\rm th}$ consumer and $k_a$ the outdegree of the $a^{\rm th}$ metabolite. We can express also this in terms of the marginal averages by defining a constant $\alpha$ which gives the relative number of metabolites to populations $M = \alpha N$ and dividing through by $N$
\begin{align}
\begin{split}
    z^{\rm in}_c = \alpha z^{\rm out}_m \\
    z^{\rm out}_c = \alpha z^{\rm in}_m.
\end{split}
\end{align}
Thus the average degrees are constrained by the relative size of the consumer and metabolite components. If they are equal $a=1$ then the average degree must be the same. 

In order to find the solutions for the proportions of consumers and metabolites present we follow a similar logic as before with one major difference. As the in- and outdegree are no longer independent we need to account for the potential correlations in the node degrees. To make this adjustment we introduce an additional step, defining the probability of arriving a node following a random outgoing link backwards and it being present as $c_1^*$ and $m_1^*$ for consumer populations and metabolites respectively. We also need to define two new generating functions which give the indegree distribution of a node we arrive at by following a outgoing link backwards. This is similar to the concept of excess degree used in the giant component literature
\begin{align}
    C_1(x) = \frac{\sum_{jk} c_{jk} x^{j} k}{\sum_{jk} c_{jk} k} = \frac{1}{z_c^{out}} \left. \frac{\delta C(x,y)}{\delta y} \right|_{y=1} .
\end{align}
We also define this for metabolites
\begin{align}
    M_1(x) = \frac{1}{z_m^{\rm out}} \left. \frac{\delta M(x,y)}{\delta y}.\right|_{y=1} 
\end{align}

Applying these to the model we arrive at a system of equations for the probability of arriving at a node and it being present $c_1^*$ and $m_1^*$ which are very similar to the uncorrelated case. 
\begin{align}
    \label{eq:si:cor_c1}
    \begin{split}
        c_1^* &= C_1(m_1^*), \\
        m_1^* &= 1 - M_1(1-c_1^*).
    \end{split}
\end{align}
The probability of arriving at a present consumer is simply the probability of arriving at a consumer with a given indegree (accounting for the correlations) and all of those requirements being present. Likewise the probability of arriving at a metabolite and it being present is the probability we arrive at a node with a given degree and then any of the production links is present. We can also write the expressions for the probability of presence for a random node in terms of the new variables as
\begin{align}
    \label{eq:si:cor_c0}
    \begin{split}
        c^* &= C(m_1^*) \\
        m^* &= 1 - M(1-c_1^*)
    \end{split}
\end{align}
which gives the full system of equations to solve. As before we can graphically solve for $c_1^*$ constructing a consistency equation from Eq.~\ref{eq:si:cor_c1}
\begin{align}
    c_1^* = C_1(1 - M(1 - c_1^*))
\end{align}
the solutions of which can be substituted into equations~\ref{eq:si:cor_c1} and \ref{eq:si:cor_c0} to get $c^*$ and $m^*$.

\subsubsection*{Random Correlated Graphs}
To illustrate the patterns of diversity in correlated cross-feeding networks we consider the case of random graphs with correlations in node in- and outdegree. Whilst there is no single canonical bivarate Poisson distribution we use a simple formulation that lets us examine the effect of the correlation. Specifically we consider the joint in- and outdegree distribution $(j,k)$ where
\begin{align}
    \label{eq:corr_pois}
    \begin{split}
        j &= j' + R \\
        k &= k' + R
    \end{split}
\end{align}
where $j'$,$k'$ and $R$ are all Poisson random variables with parameters $z^{\rm in} = z_j' - r$, $z^{\rm out} = z_k' -r$ and $r$ respectively. The dependence of the two degree distributions is determined by parameter for the shared term $r$
\begin{align}\label{eq:cov}
\begin{split}
        \text{Cov}(j,k) &= r.
\end{split}
\end{align}
The correlation is given by
\begin{align}
     \rho = \frac{r}{\sqrt{z^{\rm in} z^{\rm out}}}.
\end{align}
The value of $r$ is additionally constrained by $r < \text{min}(z^{\rm in}, z^{\rm out})$ as the rate parameters must be positive. This sets an upper bound on the correlation based on the in- and outdegree.
\begin{align}
    \label{eq:si_rho_costraint} 
    \rho < \frac{\text{min}(z^{\rm in}, z^{\rm out})}{\sqrt{z^{\rm in} z^{\rm out}}}.
\end{align}
Deriving the generating function for the bivarate Poisson is simple using the definition of a multivariate generating function 
\begin{align}
    G(x,y) &= E[x^j y^k] \\
    &= E[x^{j' + R}y^{k' + R}] \\
    &= E[x^{j'}] E[y^{k'}] E[(xy)^{R}]
\end{align}
using the fact that $j'$,$k'$ and $R$ are independent in the last step. As they are Poisson distributed each of the expectations follows the same form $F(x)=\exp(\lambda(1-x))$ giving
\begin{align}
    G(x,y) = \exp\left[z^{\rm in}(1-x) + z^{\rm out}(1-y) + \abrack{R}(1 - xy)\right]
\end{align}

we can confirm the marginal distributions still have a Poisson distribution using Eq.~\ref{eq:si:marginal}
\begin{align}
    F(x) = F(x,1) = \exp\left[z^{\rm in} + \abrack{R})(1-x)\right] 
\end{align}

\paragraph{Generating Correlated Poisson Variables}
In this section we discuss our method to generate random cross-feeding networks with degree correlations. As discussed in the material and methods, a major constraint on the degree distributions is that the number of start- and endpoints of links in the network must match for the structure to be feasible. The presence of correlations means that the links cannot be drawn independently and must be generated together.

To generate the networks we initialise the $N$ consumers and $M$ metabolites and first sample the consumer in- and out degrees from the bivariate Poisson distribution described in the previous section. We draw the degrees with desired marginal averages $z_c^{\rm in}$ and $z_c^{\rm out}$ and correlation $\rho$, with the additional constraint shown in Eq.~\ref{eq:si_rho_costraint}. 

By defining the in- and outdegree of the consumer nodes we have placed a double constraint on the resource degrees whose sum of out- and indegree must match accordingly. We can draw samples according to this constraint and a desired covariance (and thus correlation) by using the fact that a a Poisson sample conditioned on its sum has a multinomial distribution. First we write sums of the degree distributions as
\begin{align}
    \begin{split}
        S_c^{\rm in} &= S_m^{\rm out} = S_{k'} + S_{r} \\
        S_c^{\rm out} &= S_m^{\rm in} = S_{j'} + S_{r}
    \end{split}
\end{align}
where $S_c^{\rm in}$ indicates the number of links entering consumers which must equal $S_m^{\rm out}$, the number of links exiting metabolites. The last terms decompose the sums into the individual and shared components of the correlated Poisson in Eq.~\ref{eq:corr_pois}.

As we want to set the covariance externally we can obtain the distribution of $S_r$ which as the sum of $N$ Poisson variables with mean $r$ (ie.e the covariance; Eq.~\ref{eq:cov}) will follow a Poisson distribution with mean $Nr$. Thus to sample the constrained distribution we first sample the value of $S_r$, use this to solve for $S_{k'}$ and $S_{j'}$. We then sample the distributions of $k'$,$j'$ and $r$ from the correct multinomial distributions and sum them to get the individual joint degrees. 

\subsubsection*{Community Robustness, Sampling and Culture}
In this section we determine the effects of the sampling and culture of microbial communities through their cross-feeding network. We first consider how these processes are equivalent to the removal of nodes from the cross-feeding network and then derive the equations describing the state of the community after these attacks.  

As discussed in the main text the process of sampling and culturing a microbial community can be broken into two stages in which we first incompletely sample the community, keeping only a proportion $s$ of the populations. We then supply a proportion of the resources $r$ that the microbes within may require. The final community consists of the populations that were initially sampled and then able to survive on the provided resources and any created as a result of secondary metabolism. 

Considering the sampling of the community in the first step as an attack on the network is natural. When we sample we randomly miss a proportion $1-s$ of the consumer populations who are then removed from the final community. The supply of resources has a similar, though more nuanced, interpretation. When a resource is supplied to the community it is present unconditionally. This means that any consumer needing this resource will automatically have this need fulfilled and we can remove all requirement links from the metabolite. The removal of all the links is equivalent to removing the node itself. Therefore the probability of supplying a proportion of the resources $r$ is equivalent to removing them, and can be considered using the same techniques for attacks on the network.

Having established how attacks on the network are equivalent to the sampling and culture of communities we now consider how these processes affect network structure and ultimately the diversity that communities can support. It is useful to define an attack function for the node removals $A(x, a) = (1-a) + ax$ which effectively represents a coin flip in which we keep a node with probability $a$ and remove it with probability $(1-a)$. 

First we consider the consumer requirements in the final community. In the original community each consumer had needed $k$ metabolites with probability $c_k$. After the attack we remove a proportion $r$ of these needs by supplying them to the community. This is equivalent to flipping the attack coin $k$ times letting us write the final generating function for the number of requirements as
\begin{align}
    C_{\rm A}(x) = \sum c_k (r - (1-r)x)^k = C(A(x,r)).
\end{align}
The final expression can also be reached by following the dice of dice rule discussed in section~x.

Second we consider the distribution of metabolite producers in the final community. This follows the same logic as above. In the original community a metabolite has $k$ producers with probability $m_k$. Each of these $k$ consumers has a probability $s$ of being sampled in the final community. Again the total number of links in the final community is equivalent to flipping a coin $k$ times and can be written as
\begin{align}
    M_{\rm A}(x) = \sum m_k ( (1-s) - sx)^k = M(A(x,1-s)).
\end{align}
Note the attack depends on the inverse probability $1-s$ as we keep the proportion $s$ in the final community and remove the rest. In the main text we define the two attack functions directly for simplicity of notation 
\begin{align}
    S(x) &= 1-s + sx = A(x, 1-s) \\
    R(x) &= r + (1-r)x = A(x,r)
\end{align} 

With the generating functions for the structure of the network after the attack in hand it is now easy to derive the diversity in the final community after sampling and culture. The same derivation used to obtain Eq.~\ref{eq:si_sys_eq} can be used on the generating functions after the attack $C_{\rm A}(x)$ and $M_{\rm A}(x)$ giving the new diversity in the community
\begin{align}
    c^*_{\rm} &= C_{\rm A}(m^*), \\
    m^*_{\rm} &= (1 - M_{\rm A}(1 - m^*)),
\end{align}
After solving for $c^*$ and $m^*$ we apply corrections for the diversity lost and resources gained from the process directly
\begin{align}
    c^*_{\rm A} &= s c^* \\
    m^*_{\rm A} &= r^* + (1  - r^*) m^*,
\end{align}
where consumer diversity after the attack, $c^*_{\rm A}$ is scaled by the $s$ to account for the loss of species in the sampling stage and the proportion of metabolites is the sum of those supplied $r$ plus any produced by consumers (second part scaled by $1-r$).

\clearpage

\subsection*{Supplementary Figures}

\begin{figure}[h]
    \centering
    \includegraphics[width=1.0\linewidth]{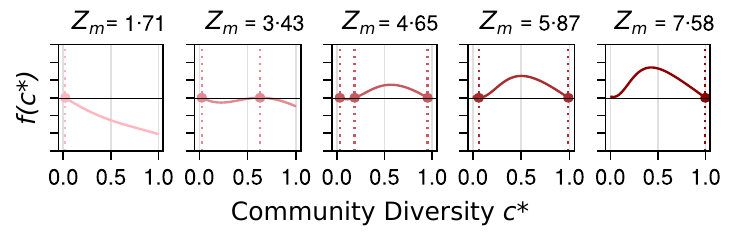}
    \caption{\textbf{Solutions to the self consistency equation} Plot showing the self consistency equation across a range of average metabolite production degree whilst holding the population requirement constant at $z_c = 4.0$. Increasing $z_m$ moving from left to right we see that below the critical point a single low-diversity solution exists. In the second panel at the critical value $z_m = z_m^c$ an additional solution appears. As $z_m$ increases three solutions are present, two of which are stable. In panel 4 we again reach a critical point where a stable and unstable equilibrium collide. After this only the high-diversity state exists. }
    \label{fig:si:graphical}
\end{figure}

\clearpage

\begin{figure}
    \centering
    \includegraphics[width=\linewidth]{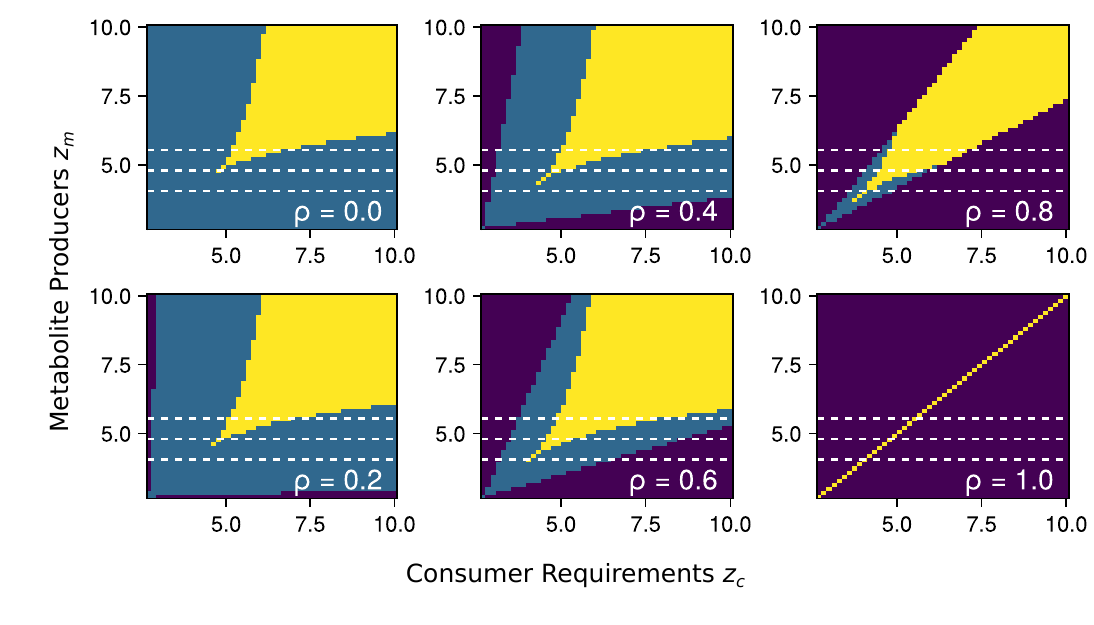}
    \caption{\textbf{Discontinuous transitions persist with degree correlations} Phase plots showing how the cusp bifurcation persists in the presence of correlations in node degree. Panels show the bifurcation surface across different levels of correlation. Yellow areas indicated the ``folded'' region where multiple steady states coexist. Blue areas have a single solution and the black areas are unfeasible (i.e. the condition on $\rho$ in Eq.~\ref{eq:si_rho_costraint} is not met). Overall increasing the correlations shifts the cusp bifurcation point to lower network degrees but also reduces the width of the bifurcation area. At high correlations the number of feasible average degree values is reduced till they must be equal $z_m = z_c$ when $\rho = 1$ }
    \label{fig:cor_phase}
\end{figure}
    
\clearpage

\begin{figure}
    \centering
    \includegraphics[width=\linewidth]{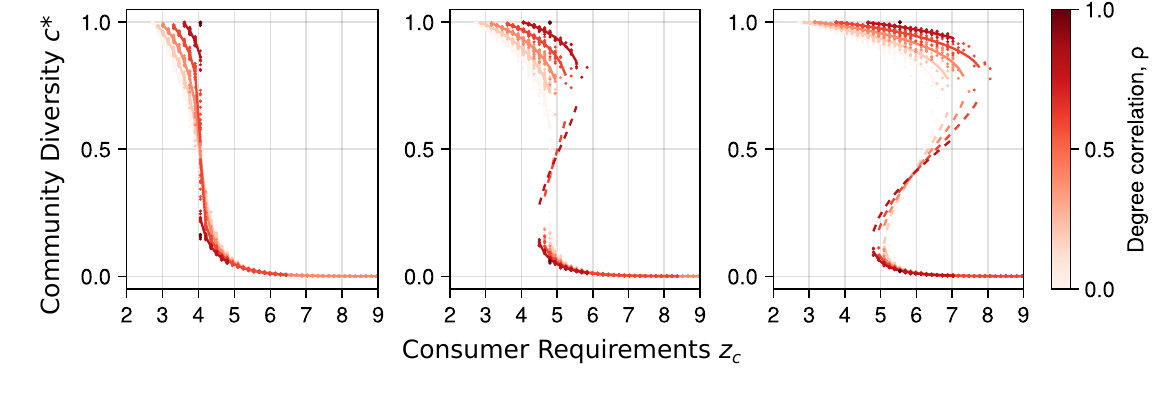}
    \caption{\textbf{The effect of degree correlations on community diversity} Plots showing how community diversity changes with the correlation between in- and outdegree in the cross-feeding network. Lines show the diversity over differing numbers of requirements $z_c$ on the x-axis and numbers of metabolite producers $z_m$ across each panel, obtained from solutions to Eqs.~\ref{eq:si:cor_c0}~\&~\ref{eq:si:cor_c1}. Colours indicate the strength of degree correlations $\rho$. Overall the analytical results match the generated networks very well. As the strength of correlations increases the region of path-dependency increases in size.}
    \label{fig:cor_div}
\end{figure}



\end{document}